# Spin-dependent scattering induced negative magnetoresistance in topological insulator $Bi_2Te_3$ nanowires

**Biplab Bhattacharyya,**[1,2] **Bahadur Singh,**[3] **R. P. Aloysius,**[1,2] **Reena Yadav,**[1,2] **Chenliang Su,**[3] **Hsin Lin,**[4] **S. Auluck,**[2] **Anurag Gupta,**[1,2] **T. D. Senguttuvan,**[1,2] **and Sudhir Husale,**[1,2] *

[1] Academy of Scientific and Innovative Research (AcSIR), National Physical Laboratory, Council of Scientific and Industrial Research, Dr. K. S Krishnan Road, New Delhi-110012, India.

[2] National Physical Laboratory, Council of Scientific and Industrial Research, Dr. K. S Krishnan Road, New Delhi-110012, India.

[3] SZU-NUS Collaborative Center and International Collaborative Laboratory of 2D Materials for Optoelectronic Science & Technology, Engineering Technology Research Center for 2D Materials Information Functional Devices and Systems of Guangdong Province, College of Optoelectronic Engineering, Shenzhen University, ShenZhen, 518060, China.

[4] Institute of Physics, Academia Sinica, Taipei, 11529, Taiwan.

*E-mail: husalesc@nplindia.org

**ABSTRACT**

Studies of negative magnetoresistance in novel materials have recently been in the forefront of spintronic research. Here, we report an experimental observation of the temperature dependent negative magnetoresistance in $Bi_2Te_3$ topological insulator (TI) nanowires at ultralow temperatures (20 mK). We find a crossover from negative to positive magnetoresistance while increasing temperature under longitudinal magnetic field. We observe a large negative magnetoresistance which reaches -22% at 8T. The interplay between negative and positive magnetoresistance can be understood in terms of the competition between dephasing and spin-orbit scattering time scales. Based on the first-principles calculations within a density functional theory framework, we demonstrate that disorder (substitutional) by Ga+ ion milling process, which is used to fabricate nanowires, induces local magnetic moments in $Bi_2Te_3$ crystal that can lead to spin-dependent scattering of surface and bulk electrons. These experimental findings show a significant advance in the nanoscale spintronics applications based on longitudinal magnetoresistance in TIs. Our experimental results of large negative longitudinal magnetoresistance in 3D TIs further indicate that axial anomaly is a universal phenomenon in generic 3D metals.

# INTRODUCTION

Over the last decade, numerous theoretical and experimental studies on topological states of quantum matter have revolutionized research in condensed matter physics. These materials possess a huge potential for tabletop experiments and advance technological applications such as topological quantum computing, spintronics, and low-dissipation electronics.[1,2] Topological insulators (TIs) are one such class of topological quantum matter that have bulk energy gap with conductive Dirac-cone-like topological surface states (TSSs) over the crystal boundary. The gapless metallic surface states of TIs are protected by time-reversal symmetry (TRS) and immune to scattering by non-magnetic impurities, thus opening new avenues for elastic scattering free transport applications. The binary Bi-based chalcogenides $Bi_2Se_3$ and $Bi_2Te_3$ have been regarded as reference 3D TIs with a relatively simple electronic structure consisting of single Dirac-cone TSS within the bulk energy gap at the centre of the Brillouin zone ($\Gamma$-point).[3,4] These materials have been extensively investigated previously for their efficient thermoelectric properties. The existence of robust and exotic TSS in $Bi_2Te_3$ has been previously demonstrated by several groups via low-temperature quantum transport measurements.[5-7] As the electrical conduction in this narrow band-gap TI material is constantly plagued by residual bulk carriers, most of the experimental studies have been performed in low-dimensional geometries such as nanowires, nanoribbons, etc. where surface contribution is dominant.

Magnetoresistance (MR) measurements provide very useful information about the type of conduction mechanism in the various materials. It is well known that conductivity corrections resulting from the quantum interference of time-reversed closed loops around an impurity or scattering point give rise to weak localization (WL) or weak antilocalization (WAL) effects, which can be directly observed through MR measurements.[8-10] WL or WAL effect is more pronounced when phase coherence length exceeds the mean free path of the quasi-particle.[9] These quantum-mechanical interference phenomena strongly depend on the temperature and are only observable at very low temperatures. In TIs, WAL effect is expected to manifest as a cusp-like feature in the MR curve at zero magnetic (B) field due to the presence of strong spin-momentum locking of the metallic TSS that leads to a $\pi$-Berry phase and suppresses backscattering.[2,8] The WAL feature has been observed in many TI materials under out-of-plane perpendicular B-field.[6,11] It was also shown in many studies that these conductivity corrections are highly influenced by the interaction of TSS and bulk carriers, thickness of the TI films, electron-electron (e-e) interaction and presence of disorder, where a crossover from WAL to WL can be observed.[12] Parallel B-field MR provides an efficient way to investigate the bulk TI states, as it is very difficult to distinguish the bulk states properties from TSS under perpendicular B-field. It has been shown that magneto-transport measurements under in-plane B-field applied parallel to the sample plane can provide an efficient way to detect bulk conductance contribution and surface-to-bulk coupling.[13,14] WL or WAL effect can also arise under parallel B-fields due to some finite thickness of the sample material, where Aharonov-Bohm (AB) phase can be acquired by the closed loop electron trajectory encircling certain amount of B-flux while undergoing inter-surface or surface-to-bulk scatterings.[14,15]

Recent demonstration of axial or chiral anomaly in Weyl semimetals via negative MR (NMR) under parallel B-field has generated tremendous interest in studying the properties of NMR in different topological materials.[16,17] Previously, NMR was observed in magnetic materials and Anderson localized 2D electron gas.[18,19] The observation of NMR under parallel B-field in materials other than topological semimetals without any Weyl nodes has led to the belief that NMR is a generic property of metallic or semiconducting materials.[20-22] Till date, several NMR observations have been reported in TIs under both perpendicular and parallel B-fields, for example, in ultrathin $Bi_2Se_3$ films due to crossover between WL and WAL of bulk electrons,[23] in ultrathin $(Bi_xSb_{1-x})_2Te_3$ films due to opening of energy gap in TSS that modifies the Berry phase and leads to competing WL and WAL at low B-fields,[24] in ultrathin $(Bi_{1-x}Sb_x)_2Te_3$ films due to disorder-induced Anderson localization,[25] in Sn-doped $Bi_2Te_3$ polycrystalline films[13] and $Bi_2(Te_{0.4}Se_{0.6})_3$ film[26] due to WL effect of bulk states, in $Bi_2Se_3$ nanoribbons due to Zeeman energy induced surface Dirac cone deformation and spin-dependent scattering of TSS,[27] in $BiSbTeSe_2$ nanoflakes due to disorder induced enhancement of defect states density from $Ar^+$ ion milling,[28] in $TlBi_{0.15}Sb_{0.85}Te_2$ due to formation of charge puddles in disordered bulk,[29] in epitaxial $Bi_2Se_3$ thin films due to underlying scattering mechanism[21] and in $Bi_2Se_{3-y}S_y$ due to non-trivial bulk conduction and appearance of electron puddles from sulphur doping.[30] Thus, there is a growing interest to understand the underlying mechanism for the origin of NMR in non-magnetic TIs.

In this study, we report the observation of NMR under in-plane B-field applied parallel to the current direction in focused-ion-beam (FIB) fabricated $Bi_2Te_3$ nanowires. Temperature dependent interplay between NMR and positive MR (PMR) has been observed in all the nanodevices. The direction of the B-field is same for entire study. Large NMR of -22% at 8 T is observed in one of the devices at 0.3 K, which is significantly higher considering TI without magnetic doping. Various physical phenomena such as competition between dephasing and spin-orbit scattering time, $Ga^+$ ion milling induced disorder and contamination leading to strong localization, electron hopping transport, charge puddle formation and spin accumulated localized regions have been discussed for possible explanation of NMR. The first-principles calculations within a density-functional theory (DFT) framework show that $Ga^+$ induced disorder (substitutional) in the $Bi_2Te_3$ crystal gives rise to a local magnetic moments in the material. Qualitative analysis of the effect of orientation of B-field, excitation current and spin-polarization of the TSS along-with the bulk conductance interference can explain the origin of NMR and other MR characteristics observed in all the devices via spin-dependent scattering mechanism of surface and bulk electrons on local magnetic moments.

## RESULTS AND DISCUSSION

**Experimental analysis**

Parallel orientation of B-field to the plane of the sample ($B_{||}$) has always been a source of some peculiar and rich physical phenomenon in TI materials, where the strong spin-momentum locking of the TSS generates a spin-polarized current under application of electric field. An in-plane B-field applied perpendicular to the current direction ($B_{||} \perp I$) perfectly aligns parallely with the spin polarization direction and when in-plane B-field is applied parallel to the current direction ($B_{||} || I$), there occurs a perpendicular orientation between B-field and spin-polarization of TSS that results into unusual MR behaviour. The same is evident from Fig. 1 in our case with $Bi_2Te_3$ nanowire devices under $B_{||} || I$ orientation. Figs. 1a, b & c depict the percentage MR change (longitudinal MR) with applied parallel B-field for devices SH-1, 2 and 5 at different temperatures. The MR change (%) can be written as

$$MR\ change\ (\%) = \frac{R(B) - R(0T)}{R(0T)} \times 100\% \qquad (1)$$

where $R(B)$ and $R(0T)$ is the resistance at any B and 0T, respectively. We observe a temperature dependent competition between the PMR and NMR in all the devices. NMR is mostly prevalent at ultralow temperatures (T < 2 K). With increasing temperature, a shifting trend from NMR to PMR is observed for all the devices. Although, sample SH5 (Fig. 1c) does not show complete PMR till 5 K, but still there starts a decreasing NMR slope since 1 K. This reflects the fact that the observed NMR phenomenon is strongly dependent on the temperature. Another significant observation in this direction is the decreasing MR change (%) with increasing temperature in all the devices. For device SH2 (Fig. 1b), a high MR change of -22% (at 8 T) is observed at ultralow temperature of 0.3 K, which gradually decreases to around 1.5% at 8 K. This decay of NMR with increasing temperature suggests the presence of some temperature dependent quantum-mechanical property related to the orientation of spin-polarization of TSS and B-field.

It must be noted that, we did not observe any periodic MR oscillations (in Fig. 1) corresponding to the cross-sectional area of the $Bi_2Te_3$ nanowire arising from AB interference effect of TSS, which has been previously shown to be a hallmark of TSS existence under $B_{||} || I$.[5,6,31] Rather, we have observed aperiodic MR fluctuations in the data, which is most likely the result of universal conductance fluctuations (UCF)[32] or quantum interference of closed loop electronic trajectories with random cross-sectional areas. With increasing temperature, a PMR starts to appear at around 1.5 K for devices SH-1, 2. For device SH5 complete PMR is expected at a higher temperature than 5 K. PMR occurs in $B_{||} || I$ orientation due to the finite thickness of the sample, where AB phases are acquired by electron wave-functions enclosing some B-flux.[14,15] The resulting quantum interference effect of self-intersecting closed paths modifies the conductivity of the sample. As we do not observe any flux quantization period corresponding to the cross-sectional area of the nanowire, therefore, we infer that high bulk conduction is present in our $Bi_2Te_3$ nanowire samples. The interference of bulk states with TSS can easily host such random electron trajectories leading to arbitrary AB phases and aperiodic MR oscillations. Previous reports on $Bi_2Te_3$ material have also demonstrated the presence of high bulk conductance contribution, where gate voltage or doping was used to suppress the bulk conductivity and realize TSS transport.[5,13]

Figure 1d depicts the competition between PMR and NMR in $Bi_2Te_3$ nanowire devices. The MR at ultralow temperature of 0.05 K can be divided into three regimes, viz., regime-1 with small NMR, regime-2 with small PMR and regime-3 with large NMR extending till very high B-fields. The different regimes for devices SH-1 and 5 range from 0 to 1 T for regime-1, 1 to 3 T for regime-2 and 3 to 10 T (which is the highest B-field used in this study) for regime-3. The main panel of Fig. 1d shows that, strangely enough, the MR for devices SH-1 and 5 enter and exit any of the regimes at around same B-field value. Also, at different temperatures (till 0.3 K) for device SH5, we observe that MR enters and exits any regime at same B-field value. But, this is not the case with device SH2, where only two regimes (regime-1 and 2) can be seen till 10 T at T = 0.05 K (inset in Fig. 1d). At 0.05 K, initially NMR is present till 2.7 T followed by PMR till 10 T. Although at 0.1 and 0.2 K, again three MR regimes were found within 10 T. For 0.1 K (inset in Fig. 1d), regimes-1, 2 and 3 extend from 0 to 1.16 T (NMR), 1.16 to 3.5 T (PMR) and 3.5 to 10 T (NMR). Similarly for 0.2 K (Fig. 1b), regimes-1, 2 and 3 extend from 0 to 0.8 T (NMR), 0.8 to 2.3 T (PMR) and 2.3 to 8 T (NMR). Thus, we conclude that the interplay between NMR and PMR is sample-specific and the magnitude of B-field to observe these effects cannot be generalized for the material. Although, high similarities between the MR trends for different samples suggests the presence of same physical phenomenon based on some common property of the $Bi_2Te_3$ material.

The temperature dependent switching of MR between NMR and PMR as seen in Fig. 1 has been previously attributed to the WL and WAL effect, respectively.[23] This is because such quantum interference effects are more likely to occur at low temperatures and eventually vanish with increasing temperature. As we observe the presence of more bulk contribution in our devices, so the WL and WAL features can be a manifestation of the quantum interference occurring in the bulk TI states. Usually, two time scales are used to describe the quantum correction to conductivity (or MR) in TIs with strong spin-momentum locking, viz., electron dephasing time ($\tau_\varphi$) and spin-orbit scattering time ($\tau_{SO}$).[23,33] For regime-1 in Fig. 1d, $\tau_{SO} > \tau_\varphi$, which means that spin-orbit scattering leading to frequent spin-flips is weak in this regime. This decreases the resistance of the material with increasing B-field and results into NMR (WL effect). Regime-2 can be described by an intermediate time scale, $\tau_\varphi > \tau_{SO}$, where spin-orbit scattering is relatively strong leading to slight increase in the material resistance, which reflects as PMR or WAL effect. This intermediate regime is manifested as a crossover from WL to WAL or vice-versa. For regime-3, $\tau_{SO} \gg \tau_{\varphi}$, which suggests very weak or negligible spin-flips leading to huge decrease in resistance. This can be observed as large NMR (WL effect) in the devices. For device SH5, regime-3 shows a large linearly decreasing MR with increasing B-field.

Previously, in ultrathin $Bi_2Se_3$ films,[23] a crossover from PMR to NMR (WAL to WL) was observed for $B_{||} \perp I$ orientation. Whereas, no such crossover was observed for $B_{||} || I$ orientation within 50 T. This is a direct consequence of the fact that for $B_{||} \perp I$, B-field and spin-polarization direction of TSS (due to applied electric field) are parallel, which leads to decreased spin-orbit scattering and high $\tau_{SO}$. In our case, we find a crossover phenomenon and huge NMR for $B_{||} || I$ in $Bi_2Te_3$ nanowires. This is obviously much unexpected because in this orientation, B-field becomes perpendicular to the spin-polarization direction of TSS. As spin-flips leading to the alignment of B-field and spin-polarization direction are hindered by the strong in-plane perpendicular spin-momentum locking of TSS, therefore, $B_{||} || I$ orientation does not decrease the spin-orbit scattering and a PMR is expected with increasing B-field. Thus, the occurrence of NMR in all the devices under this orientation is indicative of the fact that some additional physical mechanism is dominant in our samples that leads to the observed high $\tau_{SO}$.

Earlier studies have demonstrated that disorder induced localization of electronic states plays a significant role in the origin of NMR in various materials.[10] It has been theoretically predicted that under WL regime ($kl_e \gg 1$ with $k$ = wave-vector and $l_e$ = mean free path), where conduction occurs via quantum diffusion, B-field may induce a phase shift in electronic wave-function that suppresses the backscattering probabilities leading to NMR.[19,34] Whereas, under strong localization regime (Anderson localization, $kl_e < 1$), where conduction occurs via quantum jumps (variable-range-hopping (VRH)) between localized states, dephasing by B-field can reduce the interference between multiple elastic scattering paths involved in the hopping process and lead to NMR.[35] It is important to note that the scattering process involved in VRH transport is very different from conventional backscattering model in diffusive conductors. Numerical simulations have shown that NMR in VRH regime is much larger than WL regime.[36] The incoherent mechanism model by Raikh[37] predicts that the contraction of electronic wave-function in a strongly localized region under B-field can decrease the repulsive energy of neighbouring sites, which results into an increase in density of states at Fermi level and decreases resistance. In another report by Hu et al.,[38] it was shown that spatial inhomogeneities arise in silver chalcogenides due to the clustering of silver ions at grain boundaries or lattice defects leading to distorted current paths. Current jets caused by the spatial conductivity fluctuations cause longitudinal NMR in these samples. Recently, Anderson localization was observed in ultrathin $(Bi_{1-x}Sb_x)_2Te_3$ films,[25] which is a 3D TI material, where otherwise a strong immunity to

Anderson localization is predicted theoretically.[2] It was observed that a crossover from diffusive WAL to VRH transport occurs with growing disorder strength and NMR appears. A temperature dependent gigantic NMR was observed by Breunig et al.[29] in nearly bulk insulating TI material $TlBi_{0.15}Sb_{0.85}Te_2$. A disorder-related physical mechanism independent of the B-field orientation with sample was proposed, where creation of charge puddles by charged donor and acceptor atoms in imperfect compensated TIs can induce B-field-sensitive percolating current paths resulting into decreased resistance with increase in B-field. It was also shown that a crossover from NMR to PMR occurs with increasing temperature, which can be attributed to the destruction of charge puddles by thermally energetic carriers. A similar kind of electron puddle formation possibility due to sulphur doping and non-trivial bulk conduction was proposed by Singh et al.[30] in TI material $Bi_2Se_{3-y}S_y$ to explain the observed NMR. In another report, Banerjee et al.[28] demonstrated the origin of NMR due to $Ar^+$ milling process, which increases the density of defects in TI material $BiSbTeSe_2$ leading to localization of electronic states and VRH transport. It was proposed that with application of B-field, spins of the electrons trapped in localized regions align with the B-field direction resulting into decreased spin scattering and resistance, which manifests as NMR.

In the present study, we have used $Ga^+$ ion milling process to fabricate the $Bi_2Te_3$ nanowires. It is well known that ion milling process induces slight disorder in the system. There can some deformation and contamination in the material due to highly energetic Ga ions, which can create defect states. In our previous report on FIB-fabricated narrow $Bi_2Se_3$ nanowires, we had experimentally found the evidence of Efros-Shklovskii VRH mechanism, which is usually dominant in highly disordered systems.[39] This suggests the possibility of strong localization effects in this case with $Bi_2Te_3$ also, where defect states from ion milling can bring the system transport in VRH regime and spin accumulated localized regions could lead to observed NMR under B-field. Previously, in $Bi_2Se_3$ nanoribbons synthesized using chemical vapour deposition method, NMR under both $B_{||} \perp I$ and $B_{||} || I$ orientation was attributed to the Zeeman effect on TSS transport.[27] It was proposed that due to the large Landé factor ($g \sim 50$) of $Bi_2Se_3$ surface electrons, Zeeman energy from $B_{||}$ will deform the Dirac cone leading to small spin-polarization of TSS. Also, the anti-site defect in $Bi_2Se_3$ due to Bi atom replacing one $Se_2$ atom produces asymmetry in the spin-up and spin-down density of states (DOS), generating local magnetic moments in the system. So, there occurs a spin-dependent scattering of surface electrons on local magnetic moments leading to NMR.

In our current study, it is possible that under $B_{||} || I$ orientation, such spin-dependent scattering mechanism for NMR is present. As stated before, the interaction of $Ga^+$ ion with $Bi_2Te_3$ material, while fabricating $Bi_2Te_3$ nanowire from exfoliated $Bi_2Te_3$ nanoflake using FIB can introduce disorder in the system. Since, this fabrication technique is top-to-bottom approach (i.e. from nanoflake to nanowire) and milling is done from one of the sides of the nanoflake with other side remaining unaffected, therefore, we assume that most of the Ga$^+$ interaction is limited to only one side surface of Bi$_2$Te$_3$ nanowire. Bi$_2$Te$_3$ crystal structure consists of quintuple layers, i.e., five monatomic sheets of Te$_1$-Bi-Te$_2$-Bi-Te$_1$ stacked together with van der Waals (vdW) gap in between the quintuple layers (QLs).[40] The Te$_1$-Te$_1$ bond is the weakest while the Bi-Te$_1$ bond is the strongest. Therefore, mechanical exfoliation usually results into cleaving the crystal into the vdW gap and breaking of Te$_1$-Te$_1$ bond. The two types of interaction or disorder in the crystal structure of Bi$_2$Te$_3$ by Ga are possible, *viz.*, interstitial defect with Ga$^+$ ion being trapped inside the vdW gap between two QLs near to surface and substitutional defect with dislocations or line defects in the crystal structure due to Ga atom occupying a Te atom site near to surface layers. This may occur due to the Te vacancy caused by highly energetic Ga$^+$ ion, which can strike out Te atom from its site. This Ga$^+$ ion may acquire an electron and stabilize at the Te vacancy site. This can cause a line defect and dislocation in that QL. Since, the Bi-Te$_1$ bond is stronger than Bi-Te$_2$ bond;[40] therefore, we focus on Ga occupying the Te$_2$ atomic site. Thus, we expect an asymmetric spin-resolved DOS near the Fermi energy leading to local magnetic moments that induces spin-dependent scattering of bulk and surface electrons.

In order to further investigate the origin of WL-like feature at zero B-field, we perform the Altshuler and Aronov (AA) fitting of the conductance correction ($\Delta G = G(B) - G(0T)$) versus B data. Equation 2 represents the AA formula[15,41] that describes the quantum correction to conductivity under $B_\parallel \parallel I$ orientation.

$$\Delta G = G(B) - G(0T) \cong -\frac{\alpha e^2}{\pi h} ln\left[1 + \beta \left(\frac{et L_\varphi}{\hbar}\right)^2 B^2\right] \qquad (2)$$

Where $e$ = electronic charge, $h$ = Planck's constant, $\hbar = h/2\pi$, $t$ = thickness of the nanowire and $L_\varphi$ = electron dephasing length. The pre-factor $\alpha$ provides information about the strength of spin-orbit interaction (SOI) in the material with $\alpha$ = 0, 0.5 and -1 for strong magnetic scattering, WAL with strong SOI and WL with weak SOI, respectively.[14,41] The AA correction is valid for the regime where $l_e \ll t$, i.e., the dirty limit.[15] The value of $\beta$ in this regime is 0.33.[14] Figure 2 shows the AA fit for all the devices at low B-fields with $\alpha$ and $L_\varphi$ as the free parameters. The curves have been shifted for clarity, except for 1 K for SH1 (Fig. 2a), 1.5 K for SH2 (Fig. 2b) and 5 K for SH5 (Fig. 2c), which represent the true $\Delta G$ values. Figure 2d depicts the best fit $\alpha$ values. Inset in Fig. 2d shows the $L_\varphi$ values for all the devices. We observe the presence of very high $L_\varphi$ values. For all the devices, $\alpha$ value increases with temperature from an initial negative value and approaches zero. The negative $\alpha$ is a result of NMR or WL-like feature. As with increasing temperature, a crossover from NMR to PMR occurs, therefore, $\alpha$ becomes less negative and ultimately changes sign once PMR starts to appear. Some of the $\alpha$ values estimated are as follows: -0.00076 at 0.05 K and -0.00017 at 1 K for SH1, -0.1197 at 0.3 K and 0.00227 at 5 K for SH2, and -0.0179 at 0.02 K and -0.00081 at 1 K for SH5. The very small $\alpha$ values close to zero indicate the possibility of strong magnetic scattering prevailing in the system. Therefore, we carry out the first-principles analysis to verify whether or not the Ga$^+$ disorder is generating any local magnetic moments in the Bi$_2$Te$_3$ nanowire system.

**Electronic structure analysis**

To develop a better understanding of the experimental results, we have computed electronic structure of Bi$_2$Te$_3$ with Ga defects within DFT[42] framework as implemented in the Vienna Ab initio Simulation Package (VASP).[43,44] We use projector augmented wave method to treat interaction between ion cores and valence electrons and generalized gradient approximation to consider exchange-correlation effects.[44,45] All the results presented here are obtained using the fully relaxed structural parameters. The plane wave-cut off energy of 310 eV is employed and a 12 × 12 × 8 Γ-centred $k$-mesh is used for bulk computations. We use a slab model with a vacuum of 12 Å to avoid interaction between periodically repeated slabs and 9 x 9 × 1 Γ-centered $k$ mesh to obtain the surface states. In order to resolve the Dirac cone and spin structure of TSS, we carry out computations on a finer k-mesh around the surface Brillouin zone centre and obtain energy and three spin-components at each k-point from the expectation value of the three spin Pauli matrices, $\sigma_x$, $\sigma_y$, and $\sigma_z$. The effect of Ga defects is considered within supercell approach using 2 × 2 × 1 conventional hexagonal supercell with sixty atomic layers. Ga substitutional defect is modelled by replacing one Te$_2$ atom from the ideal position in pristine Bi$_2$Te$_3$ structure, whereas, for Ga interstitial defect, we place Ga atom in the van der Waals gap and perform full structural relaxation.

In order to understand the effect of Ga defects, we first study the topological properties of pristine Bi$_2$Te$_3$. The conventional bulk hexagonal unit cell is shown in Fig. 3a. It consists of 15 atomic layers that are grouped into three QLs. The associated spin-resolved density of states (DOS) is shown in Fig. 3b. The DOS of the majority and minority spin states are coloured in green and violet. We can see that the DOS of two spin states is equal, consistent with its nonmagnetic ground state with TRS. A clear band gap can be seen between the occupied valence and unoccupied conduction states. The slab band structure of Bi$_2$Te$_3$ in Fig. 3c resolves a single topological Dirac cone surface state that connects bulk valence and conduction bands. The Dirac point overlaps in energy with bulk valence continuum and lies at an energy ~ -0.1 eV. The spin-texture and Dirac cone structure of the upper portion of the TSS are shown in Fig. 3d & 3e, respectively. The spin is clearly constrained perpendicular to the momentum over a substantial region of k-space around the Dirac point. As one moves away from the Dirac point a finite out-of-the-plane spin component develops and the Dirac cone becomes hexagonal warped. The spin-momentum locking feature of the surface state spin-texture leads to the absence of backscattering over the surface. The spin-resolved DOS with Ga interstitial defect, where Ga is trapped inside the van der Waals gap (see Fig. 3f) is shown in Fig. 3g. It is found that the Ga interfacial defect does not generate any imbalance between the majority and minority spin-states. However, it uniformly electron dopes the system, thereby moving the Fermi level more into the conduction region. The substitutional defect, where Ga replaces Te$_2$, shows small imbalance in the two spin-states, indicating the existence of local magnetic moments. A careful analysis further reveals that Bi$_2$Te$_3$ QL with Ga defect possesses a local magnetic moment of 0.14 $\mu_B$ where a major contribution to magnetic moment comes from guest Ga atom. Such local magnetic moments due to substitutional defects have been reported earlier for Bi$_2$Te$_3$ and Bi$_2$Se$_3$. This magnetic moment breaks TRS and can lead to spin-dependent scattering in the system.

Figure 4 shows the schematic of the electron transport and scattering mechanism in our FIB-fabricated Bi$_2$Te$_3$ nanowires under zero B-field (Fig. 4a) and $B_{||}||$ I orientation (Fig. 4b). As confirmed by the DFT analysis (Fig. 3), Ga$^+$ disorder due to milling process does produce some local magnetic moments in the material. Also, it is expected that Ga$^+$ implantation and contamination will be at surface of the sample only, therefore, Fig. 4 depicts some Ga impurity atom with localized region of magnetic moments at surface. Excitation current (I) is applied along the nanowire length. When surface or bulk electron encounters a localized region, then the interaction between electron spin ($\vec{s_e}$) and local magnetic moment ($\vec{m}$) can be explained by the exchange energy $\beta \vec{s_e} \cdot \vec{m}$, where $\beta$ is the exchange interaction strength.[27] Figure 4a shows the scattering mechanism for B = 0 T, where the local magnetic moments are randomly oriented. Depending on the angle between $\vec{s_e}$ and $\vec{m}$, the scattering of electron occurs. For parallel $\vec{s_e}$ and $\vec{m}$, there is higher probability for electron to move forward than being scattered backwards; and for anti-parallel $\vec{s_e}$ and $\vec{m}$, backscattering probability is high. Thus, we have a spin-dependent scattering of both spin-up and spin-down electrons from both surface and bulk conduction channels in absence of B-field. With application of B-field ($B_{||}||$ I in Fig. 4b), the local magnetic moments align parallel to $\vec{B}$ along the nanowire length. But, spin-polarization of TSS is unaffected by $\vec{B}$ due to strong spin-momentum locking, resulting into $\vec{s_e} \perp \vec{m}$ for surface electrons. This means that spin scattering of TSS is highly unfavourable due to zero exchange interaction energy. Also, there is reduced spin scattering of bulk electrons as their spins align with $\vec{B}$ and local magnetic moments. Therefore, the significantly reduced spin-dependent scattering of both surface and bulk electrons from localized magnetic moments under $B_{||}||$ I leads to high $\tau_{SO}$, which decreases the sample resistance and causes NMR. The crossover from NMR to complete PMR with increasing temperature can be attributed to the diminishing density of local magnetic moments due to thermal energy.

In our previous report on FIB-fabricated Bi$_2$Se$_3$ nanowire, we found signatures of periodic AB oscillations from TSS with dominant $h/e$ flux quantization period.[46] There was no observation of NMR in any of the Bi$_2$Se$_3$ nanowires for T ≥ 2 K. However, we did find evidences of modified surface electron path due to FIB-induced disorder and Ga contamination indicating the robustness of TSS to non-magnetic disorder. Recently, few experimental reports have validated the existence of robust TSS in FIB-fabricated TI nanostructures and shown the promising nature of FIB technique towards fabrication of desired TI-based nanostructure geometries.[46-49] In this study, high bulk conductivity of $Bi_2Te_3$ and ultralow temperatures (down-to 20 mK) may have led to the observation of NMR due to enhanced localization and spin-scattering related effects on quantum transport. For regime-1, the sharp NMR cusp near 0 T can be attributed to the WL effect of bulk $Bi_2Te_3$ channels. Previous theoretical[50] and experimental[13] reports have demonstrated that WL is expected for bulk TI bands, and whenever the bulk WL channels outnumber the TSS WAL channels, NMR due to WL effect is observable. The strength of spin-orbit coupling is very strong in the bulk of 3D TI $Bi_2Te_3$ and ideally the bulk states should also demonstrate WAL effect at low fields similar to the TSS. However, unlike the gapless TSS, the gapped bulk states have relatively large bulk bandgap, which may result into WL effect. As under parallel B-field orientation, a large number of bulk channels contribute to the quantum transport, therefore the overall WL effect in regime-1 can be interpreted as the collective result of multiple bulk transport channels. The competition between the two types of scattering channels "bulk WL channels" and "bulk WAL channels" decides the overall magneto-transport behaviour of the system. Also, it is known that sample fabrication introduces some disorder and defects in the system. Thus, it becomes highly likely that the bulk WL channels will outnumber the bulk WAL channels and lead to WL cusp or NMR. In regime-2 at low B-fields, we have high spin-orbit scattering rates (low $\tau_{SO}$) due to the scattering of both spin-up and spin-down electrons from surface and bulk on some randomly oriented local magnetic moments that do not align completely parallel with the B-field direction due to low B-field strength. This scattering leads to slight increase in resistance and thus, PMR in regime-2. However, under the influence of strong B-field in regime-3, these local magnetic moments align parallel to the direction of B-field, which leads to significant decrease in the spin-dependent scattering (high $\tau_{SO}$) of surface and bulk electrons on local magnetic moments and causes resistance to decrease, i.e., NMR in regime-3.

In a recent turnover of events, it was theoretically and experimentally shown that the appearance of charge imbalance in generic 3D metals due to parallel orientation of electric and magnetic field, similar to the chiral anomaly effect in Weyl semimetals, makes axial anomaly a universal phenomenon not specific to Weyl or Dirac semimetals.[20-22,51] It was also predicted that axial anomaly does not guarantee NMR. NMR phenomenon under $B_\parallel \parallel I$ is not just dependent on the electronic band structure of a material, but also on the type of scattering mechanism in the sample. In case of ionic impurity scattering, strong NMR can be observed and in presence of both neutral and ionic impurities, a crossover from NMR to PMR can be observed under $B_\parallel \parallel I$ for any 3D or quasi-2D metal in the quantum limit.[22] Occurrence of such a longitudinal NMR has been proposed as a hallmark of bulk transport in topological phases of matter.[51] This reflects the fact that longitudinal NMR observed in 3D TIs and other non-topological materials is a condition- and sample-specific complicated phenomenon that needs many more experimental and theoretically efforts to fully explain the underlying mechanism.

**CONCLUSION**

The observation of large NMR in TI $Bi_2Te_3$ nanowires at ultralow temperatures (T < 2 K) has been reported in this work. Strong temperature dependence of NMR suggests some quantum mechanical phenomenon as the origin. MR switch to PMR with increasing temperature was observed. Similar type of MR characteristics have been observed for all the nanodevices indicating some common physical mechanism intrinsic to the material. WL and WAL-like features arising from NMR and PMR have been discussed via competing dephasing and spin-orbit scattering time scales. Different mechanisms reported to cause NMR in the past such as disorder-induced localization of electronic states, VRH transport, formation of charge puddles and ion-milling enhanced defect density in the material were discussed. Spin polarized DFT calculations confirm the presence of spin-dependent scattering of surface and bulk electrons on local magnetic moments created by $Ga^+$ disorder as the reason for observed temperature dependent NMR. A very speculative comparison with universal axial anomaly phenomenon in generic 3D or quasi-2D metal in the quantum limit under $B_{||} || I$ orientation is done. We believe that validation of universal axial anomaly in 3D TIs will require further experiments towards estimation of the quantum limit and MR characteristics at ultrahigh B-fields (~ 50 T) giving access to lowest Landau level.

**EXPERIMENTAL METHODS**

Focused-ion-beam (FIB) milling technique was used to fabricate the $Bi_2Te_3$ nanowires from micro-mechanically exfoliated thin flakes deposited on $SiO_2$/Si substrates. The substrates were pre-cleaned via chemical (acetone, iso-propanol, methanol and de-ionized water), ultrasonication and 10 min oxygen plasma treatment. Thick Au/Ti (~80/5 nm) contacts were deposited on the substrates using DC sputtering to serve as electrical contacts. $Bi_2Te_3$ bulk crystals from Alfa Aesar company were used to exfoliate thin flakes using standard scotch-tape method. The very thin nanoflakes were localized under optical microscope (Olympus MX51) and field emission scanning electron microscope (FESEM by Zeiss-Auriga). The thickness of the localized thin nanoflakes was determined via atomic force microscopy (AFM) and cross-sectional FESEM techniques. FIB milling using $Ga^+$ ions was performed to mill nanoflake into nanowire. After that, FIB-based gas injection system (GIS) was used to deposit Pt electrodes connecting pre-sputtered gold contacts and nanowire. Four-probe geometry was designed for electrical measurements. The width, thickness and channel length (distance between two voltage measuring electrodes) of the nanowires used in this study are: ~114 nm, ~ 50 nm and ~547 nm for SH1; ~302 nm, ~ 45 nm and ~ 866 nm for SH2; and ~282 nm, ~56 nm and ~ 652 nm for SH5, respectively. Insets in Fig. 1a, b and c show the FESEM images of the devices SH1, 2 and 5, respectively. The low temperature electric transport measurements were done in a dilution refrigerator (Triton 200, Oxford Instruments) with a base temperature of 10 mK and with a 14 T uniaxial magnet.

The measurement leads were incorporated with RF filters at room temperature (cut off frequency of 100 MHz) to avoid EMI due to high frequency RF radiation reaching the sample stage. A bias current of 10 nA (17Hz) derived from a lock-in-amplifier (Signal Recovery 7265) through a series resistor of 1 MΩ is used for the entire magneto-resistance measurements.

**Conflict of Interest**

The authors declare no competing financial or non-financial interests.

**Acknowledgments**

B.B. acknowledges CSIR-Senior Research Fellowship (NET). S.H. acknowledges CSIR-NPL for funding research. Authors would like to acknowledge the funding from DST, India through the project SR/S2/PU-0003/2010(G) India, for establishing the Dilution Refrigerator facility at CSIR-NPL. Also the funding from CSIR through the project AqUARIUS is acknowledged. The authors would like to thank Prof. R.C. Budhani for help and support rendered by him for establishing the Dilution Refrigerator Lab at CSIR-NPL. S.A. would like to acknowledge the use of the High Performance Computing (HPC) facilities at Physics Department of Indian Institute of Technology in Kanpur (IITK), Intra-University Accelerator Centre (IUAC) in New Delhi, Institute of Mathematical Sciences (IMSC) in Chennai, CSIR Fourth Paradigm Institute (CSIR-4PI) at Bengaluru and University of Hyderabad in Hyderabad. Work at the ShenZhen University is financially supported by the Shenzhen Peacock Plan (KQTD20160531112042971) and Science and Technology Planning Project of Guangdong Province (2016B050501005).

**Author Contributions**

S.H. planned and supervised the study. S.H. and B.B. fabricated the nanodevices. R.P.A. performed low temperature DR measurements. B.S. and S.A. performed DFT calculations, analyzed and wrote the DFT results. C.S. and H.L. assisted with the DFT analysis and results. R.Y. assisted with the literature survey and microscopy sample preparation. A.G. and T.D.S. provided FIB materials, laboratory tools and DR facility. B.B. and S.H. carried out the experimental data analysis, image graphics, manuscript writing and preparation. All authors read and commented on the manuscript.

# REFERENCES

1. Bansil, A., Lin, H. & Das, T. Colloquium: Topological band theory. *Rev. Mod. Phys.* **88**, 021004 (2016).
2. Hasan, M. Z. & Kane, C. L. Colloquium: Topological insulators. *Rev. Mod. Phys.* **82**, 3045-3067 (2010).
3. Xia, Y. *et al.* Observation of a large-gap topological-insulator class with a single Dirac cone on the surface. *Nat. Phys.* **5**, 398-402 (2009).
4. Zhang, H. *et al.* Topological insulators in $Bi_2Se_3$, $Bi_2Te_3$ and $Sb_2Te_3$ with a single Dirac cone on the surface. *Nat. Phys.* **5**, 438-442 (2009).
5. Xiu, F. *et al.* Manipulating surface states in topological insulator nanoribbons. *Nat. Nano.* **6**, 216-221 (2011).
6. Tian, M. *et al.* Dual evidence of surface Dirac states in thin cylindrical topological insulator $Bi_2Te_3$ nanowires. *Sci. Rep.* **3**, 1212 (2013).
7. Du, R. *et al.* Robustness of topological surface states against strong disorder observed in $Bi_2Te_3$ nanotubes. *Phys. Rev. B* **93**, 195402 (2016).
8. Hikami, S., Larkin, A. I. & Nagaoka, Y. Spin-Orbit Interaction and Magnetoresistance in the Two Dimensional Random System. *Prog. Theor. Phys.* **63**, 707-710 (1980).
9. Evers, F. & Mirlin, A. D. Anderson transitions. *Rev. Mod. Phys.* **80**, 1355-1417 (2008).
10. Lee, P. A. & Ramakrishnan, T. V. Disordered electronic systems. *Rev. Mod. Phys.* **57**, 287-337 (1985).
11. Ando, Y. Topological Insulator Materials. *J. Phys. Soc. Jpn.* **82**, 102001 (2013).
12. Liu, W. E., Hankiewicz, E. M. & Culcer, D. Weak Localization and Antilocalization in Topological Materials with Impurity Spin-Orbit Interactions. *Materials* **10** (2017).
13. Zhang, H. B. *et al.* Weak localization bulk state in a topological insulator $Bi_2Te_3$ film. *Phys. Rev. B* **86**, 075102 (2012).
14. Lin, C. J. *et al.* Parallel field magnetoresistance in topological insulator thin films. *Phys. Rev. B* **88**, 041307 (2013).
15. B. L. Al'tshuler & Aronov, A. G. Magnetoresistance of thin films and of wires in a longitudinal magnetic field. *JETP Lett.* **33**, 499 (1981).
16. Kim, H.-J. *et al.* Dirac versus Weyl Fermions in Topological Insulators: Adler-Bell-Jackiw Anomaly in Transport Phenomena. *Phys. Rev. Lett.* **111**, 246603 (2013).
17. Son, D. T. & Spivak, B. Z. Chiral anomaly and classical negative magnetoresistance of Weyl metals. *Phys. Rev. B* **88**, 104412 (2013).
18. Ramirez, A. P. Colossal magnetoresistance. *J. Phys. Condens. Mat.* **9**, 8171 (1997).
19. Jiang, H. W., Johnson, C. E. & Wang, K. L. Giant negative magnetoresistance of a degenerate two-dimensional electron gas in the variable-range-hopping regime. *Phys. Rev. B* **46**, 12830-12833 (1992).
20. Kikugawa, N. *et al.* Interplanar coupling-dependent magnetoresistivity in high-purity layered metals. *Nat. Commun.* **7**, 10903 (2016).
21. Wiedmann, S. *et al.* Anisotropic and strong negative magnetoresistance in the three-dimensional topological insulator $Bi_2Se_3$. *Phys. Rev. B* **94**, 081302 (2016).
22. Goswami, P., Pixley, J. H. & Das Sarma, S. Axial anomaly and longitudinal magnetoresistance of a generic three-dimensional metal. *Phys. Rev. B* **92**, 075205 (2015).

23  Wang, H. *et al.* Crossover between Weak Antilocalization and Weak Localization of Bulk States in Ultrathin $Bi_2Se_3$ Films. *Sci. Rep.* **4**, 5817 (2014).

24  Lang, M. *et al.* Competing Weak Localization and Weak Antilocalization in Ultrathin Topological Insulators. *Nano Lett.* **13**, 48-53 (2013).

25  Liao, J. *et al.* Observation of Anderson Localization in Ultrathin Films of Three-Dimensional Topological Insulators. *Phys. Rev. Lett.* **114**, 216601 (2015).

26  Wang, Z., Wei, L., Li, M., Zhang, Z. & Gao, X. P. A. Magnetic Field Modulated Weak Localization and Antilocalization State in $Bi_2(Te_xSe_{1-x})_3$ Films. *phys. status solidi (b)* **255**, 1800272 (2018).

27  Wang, L.-X. *et al.* Zeeman effect on surface electron transport in topological insulator $Bi_2Se_3$ nanoribbons. *Nanoscale* **7**, 16687-16694 (2015).

28  Banerjee, K. *et al.* Defect-induced negative magnetoresistance and surface state robustness in the topological insulator $BiSbTeSe_2$. *Phys. Rev. B* **90**, 235427 (2014).

29  Breunig, O. *et al.* Gigantic negative magnetoresistance in the bulk of a disordered topological insulator. *Nat. Commun.* **8**, 15545 (2017).

30  Singh, R. *et al.* Unusual negative magnetoresistance in $Bi_2Se_{3-y}S_y$ topological insulator under perpendicular magnetic field. *Appl. Phys. Lett.* **112**, 102401 (2018).

31  Peng, H. *et al.* Aharonov-Bohm interference in topological insulator nanoribbons. *Nat. Mater.* **9**, 225-229 (2010).

32  Lee, P. A. & Stone, A. D. Universal Conductance Fluctuations in Metals. *Phys. Rev. Lett.* **55**, 1622-1625 (1985).

33  Bergmann, G. Weak localization in thin films: a time-of-flight experiment with conduction electrons. *Phys. Rep.* **107**, 1-58 (1984).

34  Bergmann, G. Physical interpretation of weak localization: A time-of-flight experiment with conduction electrons. *Phys. Rev. B* **28**, 2914-2920 (1983).

35  V. L. Nguen, B. Z. Spivak & Shklovskii, B. I. Aaronov-Bohm oscillations with normal and superconducting flux quanta in hopping conductivity. *Pis'ma Zh. Eksp. Teor. Fiz.* **41**, 35-38 (1985).

36  Zhao, H. L., Spivak, B. Z., Gelfand, M. P. & Feng, S. Negative magnetoresistance in variable-range-hopping conduction. *Phys. Rev. B* **44**, 10760-10767 (1991).

37  Raikh, M. E. Incoherent mechanism of negative magnetoresistance in the variable- range-hopping regime. *Solid State Commun.* **75**, 935-938 (1990).

38  Hu, J., Rosenbaum, T. F. & Betts, J. B. Current Jets, Disorder, and Linear Magnetoresistance in the Silver Chalcogenides. *Phys. Rev. Lett.* **95**, 186603 (2005).

39  Bhattacharyya, B. *et al.* Evidence of robust 2D transport and Efros-Shklovskii variable range hopping in disordered topological insulator ($Bi_2Se_3$) nanowires. *Sci. Rep.* **7**, 7825 (2017).

40  Teweldebrhan, D., Goyal, V. & Balandin, A. A. Exfoliation and Characterization of Bismuth Telluride Atomic Quintuples and Quasi-Two-Dimensional Crystals. *Nano Lett.* **10**, 1209-1218 (2010).

41  Zhao, B. *et al.* Weak antilocalization in $Cd_3As_2$ thin films. *Sci. Rep.* **6**, 22377 (2016).

42  Hohenberg, P. & Kohn, W. Inhomogeneous Electron Gas. *Phys. Rev.* **136**, B864-B871 (1964).

43  Kresse, G. & Furthmüller, J. Efficient iterative schemes for ab initio total-energy calculations using a plane-wave basis set. *Phys. Rev. B* **54**, 11169-11186 (1996).

44  Kresse, G. & Joubert, D. From ultrasoft pseudopotentials to the projector augmented-wave method. *Phys. Rev. B* **59**, 1758-1775 (1999).

45  Perdew, J. P., Burke, K. & Ernzerhof, M. Generalized Gradient Approximation Made Simple. *Phys. Rev. Lett.* **77**, 3865-3868 (1996).

46  Bhattacharyya, B. *et al.* Observation of quantum oscillations in FIB fabricated nanowires of topological insulator ($Bi_2Se_3$). *J. Phys. Condens. Mat.* **29**, 115602 (2017).

47  Bhattacharyya, B., Sharma, A., Awana, V. P. S., Senguttuvan, T. D. & Husale, S. FIB synthesis of $Bi_2Se_3$ 1D nanowires demonstrating the co-existence of Shubnikov–de Haas oscillations and linear magnetoresistance. *J. Phys. Condens. Mat.* **29**, 07LT01 (2017).

48  Friedensen, S., Mlack, J. T. & Drndić, M. Materials analysis and focused ion beam nanofabrication of topological insulator $Bi_2Se_3$. *Sci. Rep.* **7**, 13466 (2017).

49  Frolov, A. V. *et al.* Magneto-quantum oscillations in $Bi_2Se_3$ nanowires. *J. Phys. Conf. Ser.* **941**, 012063 (2017).

50  Lu, H.-Z. & Shen, S.-Q. Weak localization of bulk channels in topological insulator thin films. *Phys. Rev. B* **84**, 125138 (2011).

51  Assaf, B. A. *et al.* Negative Longitudinal Magnetoresistance from the Anomalous N=0 Landau Level in Topological Materials. *Phys. Rev. Lett.* **119**, 106602 (2017).

**Figures**

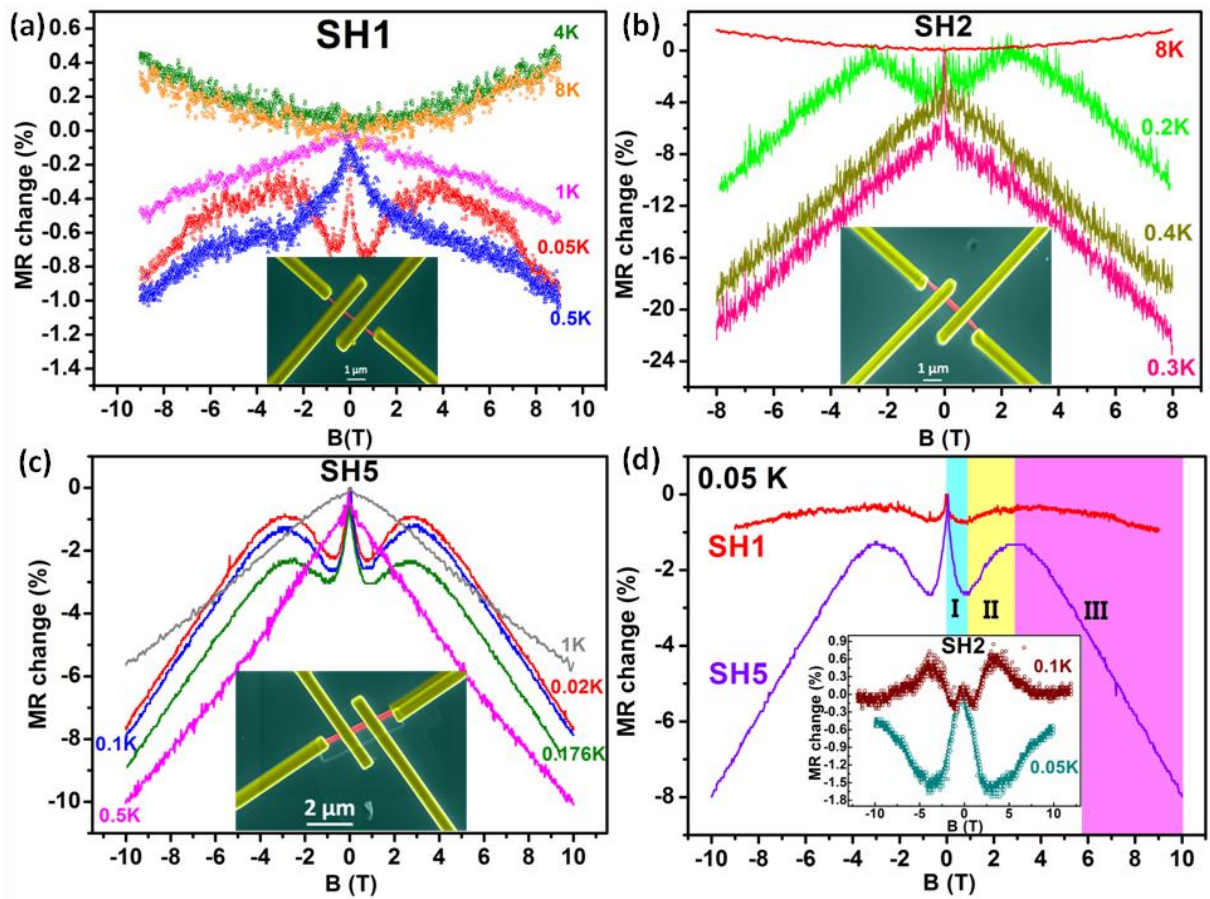

**Fig. 1. MR data for devices SH1, 2 and 5 under $B_{||} || I$ orientation.** (a, b) Temperature dependent MR switch from negative to positive for devices SH1 and 2 shown by MR change (%) curves. Insets show the false coloured FESEM images of the devices. High NMR of around 20 % has been observed for device SH2. (c) NMR for device SH5. PMR was not observed but a decrease in NMR slope suggests PMR at higher temperatures. Inset depicts the false coloured FESEM image of SH5. (d) Three different regimes depicting the interplay between NMR and PMR in all three devices at 0.05 K. Inset shows the MR for device SH2 at 0.05 K and 0.1 K. No curve shifting has been performed in any of the graphs.

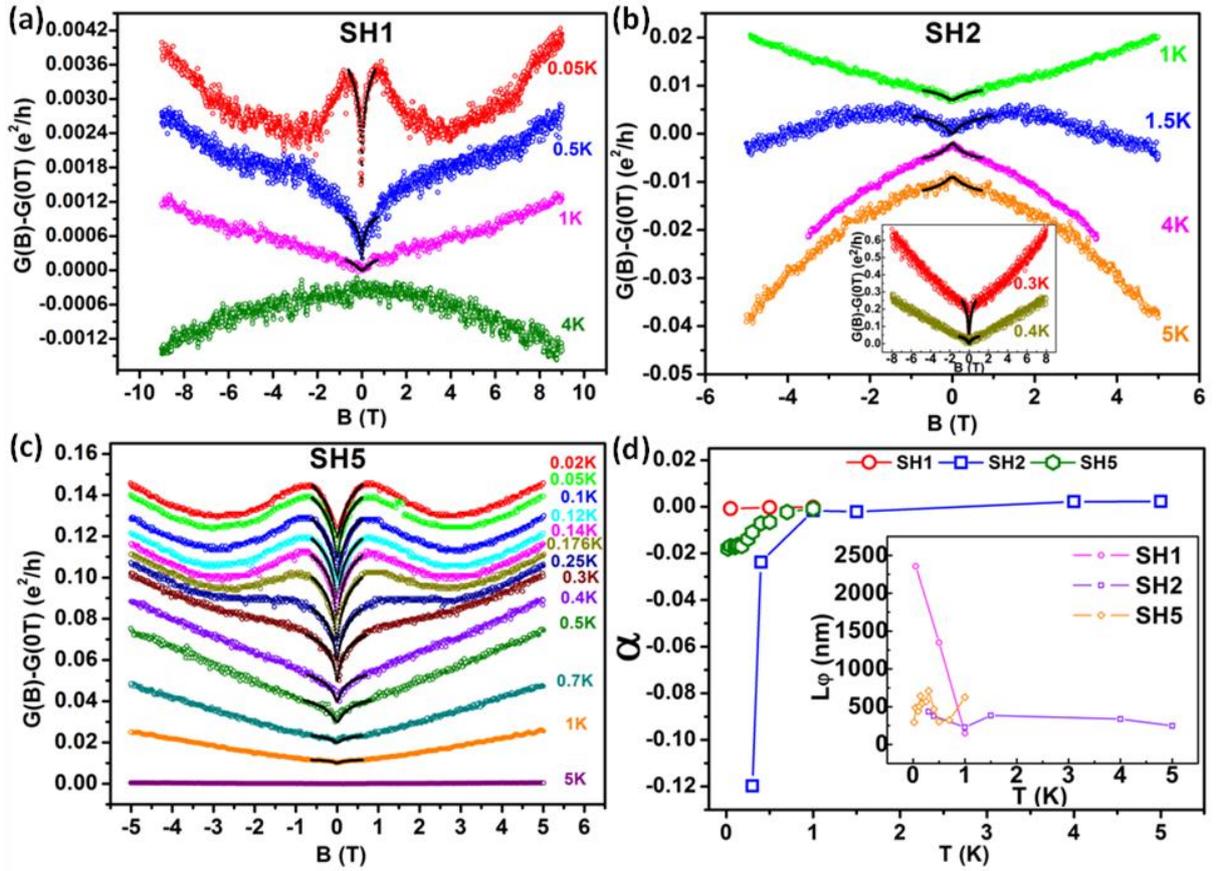

**Fig. 2. Altshuler and Aronov (AA) fitting of the conductance correction ($\Delta G = G(B) - G(0T)$) versus B data.** (a, b and c) AA fitting at low B-field MR data for the devices SH1, 2 and 5, respectively, is shown. AA fitting is shown by black dotted lines super-imposed on the WL-like dip at B = 0 T in the conductance change data. The curves have been shifted for clarity, except for 1 K for SH1 in (a), 1.5 K for SH2 in (b), 0.4 K for SH2 in inset of (b) and 5 K for SH5 in (c), which represent the true $\Delta G$ values. (d) Best fit $\alpha$ and $L_\varphi$ (inset) values for all the devices. Very small negative $\alpha$ values with high $L_\varphi$ values were estimated.

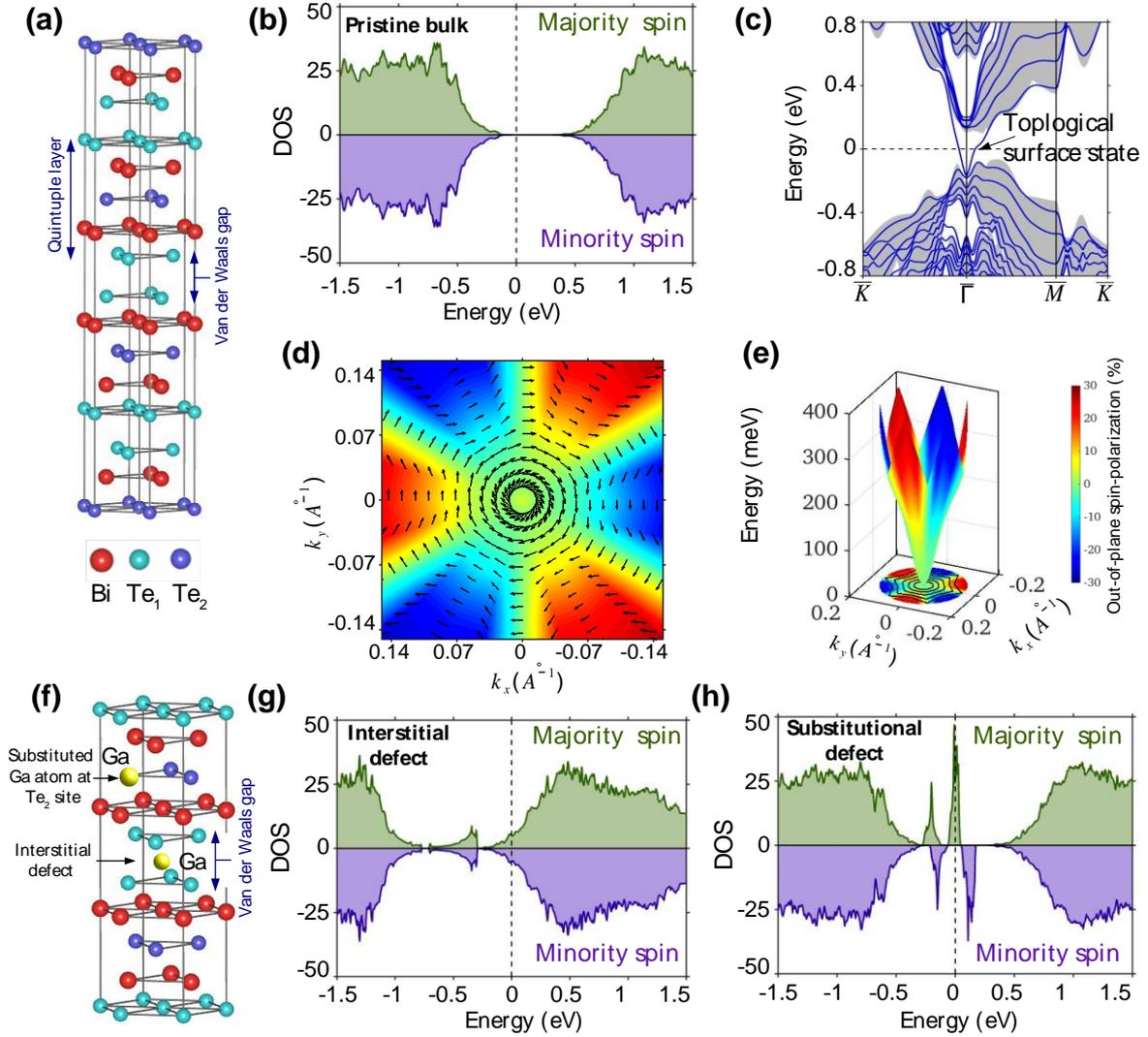

**Fig. 3. Electronic properties of Bi$_2$Te$_3$ with Ga defects.** (a) Conventional hexagonal unit cell with 15 atomic layers that are grouped in three QLs. Each QL has Te$_1$-Bi-Te$_2$-Bi-Te$_1$ atomic stacking with strong bonds between the layers whereas two QLs are held together by weak Van der Waals forces. (b) Total spin-resolved bulk DOS of $2 \times 2 \times 1$ supercell of pristine Bi$_2$Te$_3$ with a clear band gap between the occupied valence and unoccupied conduction states. (c) Calculated slab band structure of Bi$_2$Te$_3$. Shaded background grey region highlights projected bulk bands and blue thick lines identify slab bands. The TSS are clearly resolved within the bulk energy gap. (d) Helical spin-texture and (e) Dirac cone electronic structure of the top cone of TSS. In-plane spin-texture is shown with black arrows in (d). (f) Bi$_2$Te$_3$ QLs with Ga interstitial and substitutional defects. Ga atoms slide into the van der Waals gap in interstitial defects whereas these replace Te$_2$ atoms in substitutional defects. Total spin-resolved $2 \times 2 \times 1$ bulk DOS for (g) interstitial defect and (h) substitutional defect.

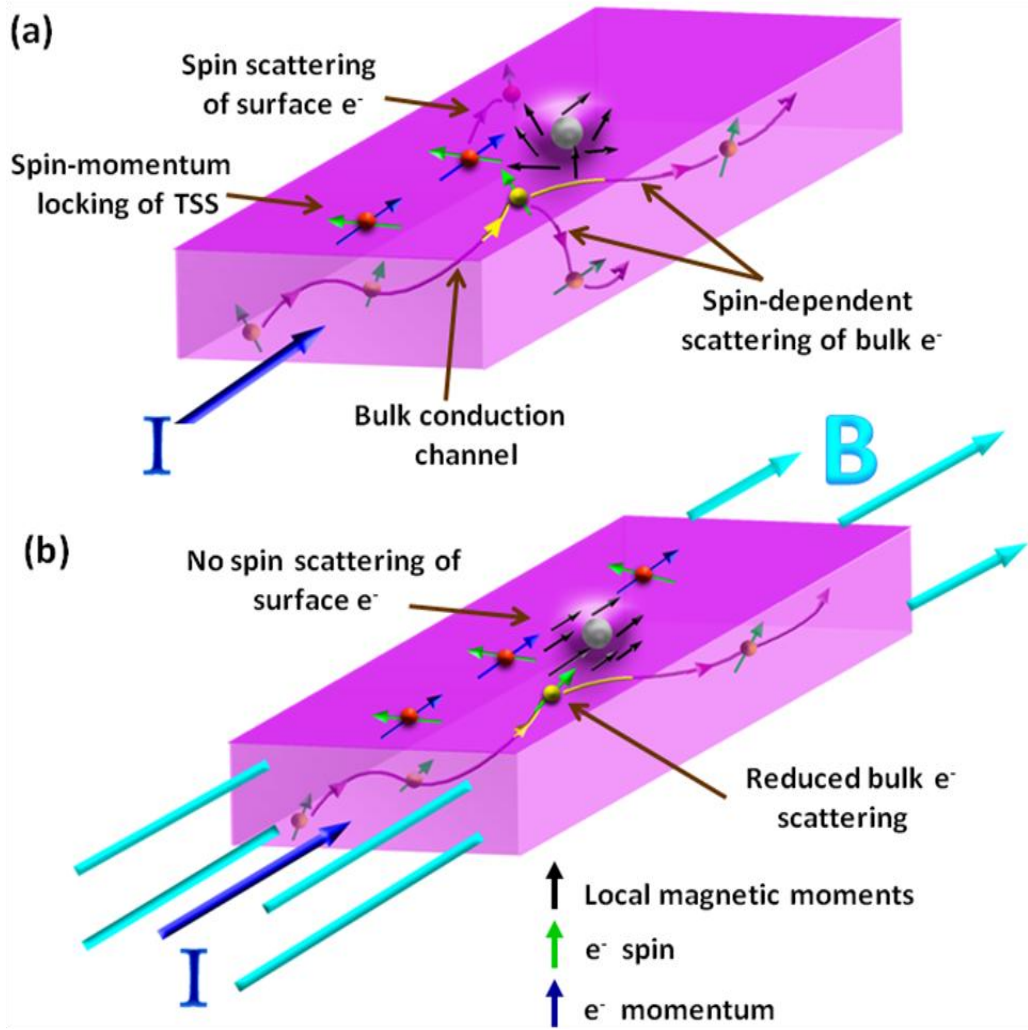

**Fig. 4. Schematic representation of spin-dependent scattering in our case with Ga disorder induced local magnetic moments.** (a) Transport under B = 0 T with applied current I. Local magnetic moments (black arrows) around Ga impurity induced disorder (grey sphere and nearby region on the surface) are oriented randomly and therefore scatter both spin-up and spin-down electrons from surface (red sphere) and bulk (yellow sphere). Strong spin-momentum locking of TSS is shown via in-plane perpendicular orientation of surface electron spin (green arrow) with surface electron momentum (indigo arrow). High bulk conduction and spin-dependent scattering on local magnetic moments is shown via transport path of yellow sphere. Yellow colour becomes evident when bulk e⁻ comes to surface. (b) Transport under $B_{||}$|| I with applied current I. Alignment of local magnetic moments in the direction of B-field reduces scattering in the system.